\documentclass{mn2e}
\usepackage{times}

\newcommand{\beq}{\begin{equation}}
\newcommand{\eeq}{\end{equation}}
\newcommand{\lab}{\label}

\newcommand{\bfxi}{\mbox{\boldmath $\xi$}}
\newcommand{\bfom}{\mbox{\boldmath $\omega$}}

\newcommand{\bfx}{ {\bf x} }

\begin{document}

\title{Detecting gravito-magnetism with rotation of polarization by a
gravitational lens}

\author[M. Sereno]
{M. Sereno$^{1,2,3}$\thanks{E-mail: Mauro.Sereno@na.infn.it}
\\
$^{1}$Istituto Nazionale di Astrofisica - Osservatorio Astronomico di
Capodimonte, Salita Moiariello, 16, 80131 Napoli, Italia
\\
$^{2}$Dipartimento di Scienze Fisiche, Universit\`{a} degli Studi di
Napoli ``Federico II", Via Cinthia, Monte S. Angelo, \\ 80126 Napoli,
Italia
\\
$^{3}$Istituto Nazionale di Fisica Nucleare, Sez. Napoli, Via Cinthia,
Monte S. Angelo, 80126 Napoli, Italia}

\maketitle

\begin{abstract}
We discuss the effects of an isolated gravitational lens on the
rotation of the plane of polarization of linearly polarized light
rays, the so called gravitational Faraday rotation, in metric theories
of gravity. By applying the thin lens approximation, we derive simple
expressions for extended, rotating astrophysical systems deflecting
electromagnetic radiation from background sources. Higher order
corrections and the case of a Kerr black hole are also considered to
show how the rotation of polarization is a purely gravito-magnetic
effect. Prospects of a future detection of a gravitational Faraday
rotation are examined.
\end{abstract}


\section{Introduction}

Conceivable post-Newtonian theories of gravity, such as general
relativity, predict space-time curvature by mass-energy currents
relative to other masses. This feature is known as intrinsic
gravito-magnetism \cite{ci+wh95}.

The occurrence of several phenomena is due to gravito-magnetism. Small
test-gyroscopes rotate with respect to distant stars in the vicinity
of a spinning body due to its rotation. This is called the
Lense-Thirring precession. Together with test particles and
gyroscopes, also photons and clocks probe some gravito-magnetic
phenomena. Bending and time delay of electromagnetic waves are two
important phenomena predicted by theories of gravity. Gravito-magnetic
effects due to the spin of an astrophysical object might be detected
in gravitational lensing events \cite{io02frm,se+ca02}. The rotation
of the lens induces particular signatures on both the position and
magnification of multiple images of a background source, also
perturbing the microlensing induced amplification curve
\cite{io03hom}.

Another well-known gravitational effect due to space-time curvature by
mass currents is the rotation of the plane of polarization for
linearly polarized light rays. Under geometric optics, a light ray
follows a null geodesic regardless of its polarization state and the
polarization vector is parallel transported along the ray
\cite{mtw73}. Such a gravitational rotation of the plane of
polarization in stationary space-times is a gravitational analogue of
the electro-magnetic Faraday effect \cite{pi+sa85,ish+al88,nou99},
i.e., the rotation that a light ray undergoes when passing through
plasma in the presence of a magnetic field. The gravitational effect
is known as gravitational Faraday rotation or Rytov or Skrotskii
effect.

From the original investigations in Skrotskii~\shortcite{skr57} and
Rytov~\shortcite{ryt38}, the effect of a gravitational field on the
plane of polarization of electromagnetic waves has been addressed by
several authors
\cite{bal58,ple60,god70,su+ma80,con+al80,pi+sa85,ish+al88,dy+sh92,nou99,ko+ma02,io04far}.
The case of a weak gravitational field has deserved particular
attention. In Plebanski~\shortcite{ple60}, the Maxwell's equations in
the gravitational field of an isolated physical system were solved and
the rotation of the plane of polarization around the propagation
vector was considered. In Kopeikin \& Mashoon~\shortcite{ko+ma02}, a
formula describing the Skrotskii effect for arbitrary translational
and rotational motion of gravitating bodies was derived by solving the
equations of motion of a light ray in the first post-Minkowskian
approximation. In Sereno~\shortcite{io04far}, the general formula for
the angle of rotation of the plane of polarization of a linearly
polarized electromagnetic wave in a stationary space-time was
re-obtained with an heuristic approach based on Mach's principle on
dragging of inertial frames, without integrating the equations of
motion.

The assumption of the existence of a dynamical space-time curvature,
as opposed to a flat space-time of special relativity, is shared among
different viable theories of gravity, such as general relativity,
Brans-Dicke theory and the Rosen bimetric theory. Metric theories of
gravity are defined such that \cite{wil93,ci+wh95}: {\it i)}
space-time is a Lorentzian manifold; {\it ii)} the world lines of test
bodies are geodesics; {\it iii)} the equivalence principle holds in
the medium strong form. The usual rules for the motion of particles
and photons in a given metric still apply, but the metric may be
different from that derived from the Einstein's field equations.

A comparison among general relativity and other viable theories of
gravity can be led with suitable tests on the basis of higher-order
effects, such as intrinsic gravito-magnetism. To date, results from
laser-ranged satellites provide the only demonstration of the
gravito-magnetic field. In 1995-2002, the Lense-Thirring precession,
due to the Earth spin, was measured by studying the orbital
perturbations of LAGEOS and LAGEOS II satellites \cite{ciu+al98}. Its
experimental value agrees with about 20\% accuracy with the prediction
of general relativity. The NASA's Gravity Probe-B satellite should
improve this measurement to an accuracy of 1\%.

In this paper, we discuss the gravitational Faraday rotation by an
isolated rotating gravitational deflector in the standard framework of
gravitational lensing \cite{pet+al01,sef}, when the source of
radiation and the observer are remote from the source of the
gravitational field (lens). As known,thanks to some usual
approximations, the effect of the spin of the deflector on the lensing
potential can be treated in a quite simple way \cite{io02frm}. The
lens equation can be modified to consider how some observable
quantities, like position and amplification of images, are perturbed
by the rotation \cite{se+ca02,io03hom}. Here, we want to extend such a
formalism to the Skrotskii effect and to consider some extended lenses
of astrophysical interest. Standard assumptions allow us to compute
higher-order approximation terms in the calculation, so that a general
treatment of the Skrotskii effect can be performed in metric theories
of gravity.

The paper is as follows. In Sec.~\ref{weak}, the weak-field, slow
motion approximation in metric theories of gravity is introduced and
the weak field limit of the gravitational Faraday rotation is
performed. In Section~\ref{thin}, we derive some simple formulae,
using the thin lens approximation, to treat systems of astrophysical
interest. Higher order correction to the Skrotskii effect are
discussed in Sec.~\ref{high}; then, the case of light rays propagating
in the vacuum region outside a Kerr black hole is considered.
Section~\ref{conc} contains some final considerations.

\section{The weak field limit}
\label{weak}

Standard hypotheses of gravitational lensing \cite{pet+al01,sef}
assume that the gravitational lens is localized in a very small region
of the sky and its lensing effect is weak. In
Sereno~\shortcite{io03ppn}, an approximate metric element generated by
an isolated mass distribution was written in the weak field regime and
slow motion approximation, up to the post-post-Newtonian order, and
with non-diagonal components accounting for effects of gravity by
currents of mass. In an asymptotically Cartesian coordinate system,
the metric $g_{\alpha
\beta}$\footnote{Latin indices run from 1 to 3, whereas Greek indices
run from 0 to 3.} can be expressed as
\begin{equation}
\lab{wf1}
ds^2 \simeq \left( 1+2\frac{\phi}{c^2} + {\cal{O}}(\varepsilon^4)
\right)c^2dt^2-
\left( 1-2\gamma \frac{\phi}{c^2} + {\cal{O}}(\varepsilon^4) \right)
\delta_{ij} d x^{i}d x^{j}
-\frac{8\mu}{c^3} \left( V_i dx^i \right)\left( cdt\right ),
\end{equation}
where $\varepsilon \ll 1$ denotes the order of approximation. In the
above metric element, $\gamma$ is a standard coefficient of the
post-Newtonian parametrized expansion of the metric tensor
\cite{ci+wh95,wil93}, measuring space curvature produced by mass. In
general relativity, it is $\gamma =1$; in the Brans-Dicke theory,
$\gamma = \frac{1+\omega}{2+\omega}$. $\mu$ is a non standard
parameter which quantifies the contribution to the space-time
curvature of the mass-energy currents. It measures the strength of the
intrinsic gravito-magnetic field \cite{ci+wh95}. In general
relativity, $\mu =1$; in Newtonian theory, $\mu =0$. Additional terms
in a parametrized expansion of the metric element can also be
considered \cite{ma+ri81,ri+ma82a}. Preferred frame effects,
violations of conservation of four momentum, preferred location
effects are not considered in our approximate metric element. In
Eq.~(\ref{wf1}), $\phi$ is the Newtonian potential,
\begin{equation}
\lab{wf2}
\phi (t, \bfx) \simeq -G \int_{\Re^3} \frac{\rho (t, \bfx^{'})}{ | \bfx -
\bfx^{'}|} d^3 x^{'},
\end{equation}
where $G$ is the gravitational Newton constant and $\rho$ is the mass
density of the deflector; $\phi/c^2$ is order $\sim {\cal
O}(\varepsilon^2)$.

${\bf V} \left(= \stackrel{3}{\bf V} + \stackrel{5}{\bf V} + {\cal
O}(\varepsilon^7)\right)$ is a vector potential taking into account
the gravito-magnetic field produced by mass currents. To the lowest
order of approximation,
\begin{equation}
\lab{wf3}
\stackrel{3}{V_i} (t, \bfx) \simeq -G \int_{\Re^3} \frac{(\rho v^i)(t, \bfx^{'})}{ |
\bfx -\bfx^{'}|} d^3 x^{'},
\end{equation}
where ${\bf v}$ is the velocity field of the mass elements of the
deflector\footnote{Bold symbols denote spatial vectors. The
controvariant components of spatial three-vectors are equal to the
spatial components of the corresponding four-vectors.}.

We assume that, during the time light rays take to traverse the lens,
the potentials in Eqs.~(\ref{wf2},~\ref{wf3}) vary negligibly little.
Then, the lens can be treated as stationary. In
Eqs.~(\ref{wf2},~\ref{wf3}), we have neglected the retardation
\cite{sef}.

The polarization vector is dragged along by the rotation of the
inertial frames. Once we have introduced the notation
\begin{equation}
\label{far2}
h \equiv g_{00}, \ A_i \equiv -\frac{g_{0i}}{g_{00}},
\end{equation}
the net angle of rotation around the unit tangent three-vector
$\hat{\bf k}$ along the path between the source and the observer, in a
stationary space-time, reads \cite{nou99,io04far}
\begin{equation}
\label{far3}
\Omega_{\rm Sk} = - \frac{1}{2} \int_{\rm sou}^{\rm obs}
\sqrt{h} \nabla {\times}{\bf A} {\cdot} {\bf d x},
\end{equation}
where ${\bf d x}=\hat{\bf k}dl_{\rm P} $ and $d l_{\rm P}$ is the
spatial distance in terms of the spatial metric $\gamma_{ij}$
\cite{la+li85},
\begin{equation}
d l_{\rm P}^2 \equiv \left( -g_{ij}
+\frac{g_{0i}g_{0j}}{g_{00}}\right)dx^i d x^j \equiv \gamma_{i j} dx^i
d x^j.
\end{equation}
Operations on three-vectors are defined in the three-dimensional space
with metric $\gamma_{\alpha \beta}$.

In Dyer \& Shiver~\shortcite{dy+sh92}, it was shown, using symmetry
arguments in a relativistic context, that, for astrophysically
interesting cases of both observer and source at large distances from
the lens, a nonshperical, nonrotating lens cannot cause a net rotation
of the polarization vector. This argument is confirmed by
Eq.~(\ref{far3}): the Skrotskii effect is due to mass-currents, i.e.,
it is a purely gravito-magnetic phenomenon.

As can be seen from Eq.~(\ref{far3}), the order of approximation is
determined by off-diagonal components of the metric. The main term is
order $\sim {\cal{O}}(\varepsilon^3)$. To determine a gravito-magnetic
effect to the second leading order, i.e., to
${\cal{O}}(\varepsilon^5)$, we need to consider terms in $g_{00}$ and
$g_{ii}$ up to ${\cal{O}}(\varepsilon^2)$. Higher order terms would
give contributions $\sim {\cal{O}}(\varepsilon^4) {\times}
\stackrel{3}{V} \sim {\cal{O}}(\varepsilon^7)$ to the Faraday rotation.
$\Omega_{\rm Sk}$ is at best $\sim {\cal{O}}(\varepsilon^3)$ unlike
the bending of light, which is $\sim {\cal{O}}(\varepsilon^2)$. That
is why the gravitational Faraday rotation is usually neglected in
light propagation analyses.

In the weak field limit, $h$ and ${\bf A}$ are simply related to the
gravitational potentials. It is
\begin{eqnarray}
h & \simeq & 1+2 \frac{\phi}{c^2} + {\cal{O}}(\varepsilon^4),
\label{far5} \\
A_i & \simeq &  \frac{4 \mu}{c^3}\left( \stackrel{3}{V_i} +
\stackrel{5}{V_i} - 2\stackrel{3}{V_i}
\frac{\phi}{c^2} \right) + {\cal{O}}(\varepsilon^7) \label{far6};
\end{eqnarray}
the proper arc length reads
\begin{equation}
\label{delppN1}
d l_{\rm P} \simeq \left\{ 1- \gamma \frac{\phi}{c^2} + {\cal O}
(\varepsilon^3) \right\} d l_{\rm E},
\end{equation}
where $d l_{\rm E} \equiv \sqrt{ \delta_{ij}d x^i d x^j}$ is the
Euclidean arc length, whereas the spatial metric reduces to
\begin{equation}
\gamma_{ij}= \left( 1-2 \frac{\phi}{c^2} + {\cal O}
(\varepsilon^4) \right)\delta_{ij} ,
\end{equation}

To calculate the gravitational Faraday rotation to order $G^{N}$, we
need the path of the deflected light ray to the order $G^{N-1}$
\cite{fi+fr80,io03ppn}. Taking the $x^3$-axis along the unperturbed
photon path, to the first order in $G$, the unit propagation vector
reads ${\bf
\hat{k}}_{(1)}
\simeq
\left\{ {\cal O} (\varepsilon^2), {\cal O} (\varepsilon^2), 1+
\gamma \phi , \right\}$. The Faraday rotation to order ${\cal O} (\varepsilon^5)$
turns out
\begin{eqnarray}
\Omega_{\rm Sk} & \simeq & -\frac{2\mu}{c^3} \left\{ \int_{\rm l.o.s.}
\stackrel{3}{V}_{2,1}-\stackrel{3}{V}_{1,2} d l_{\rm E} \label{far8}
\right. \\
& + &  \left[ \int_{\rm l.o.s.}
\left\{ \stackrel{3}{V}_{2,1}-\stackrel{3}{V}_{1,2} \right\} d l_{\rm E}
- \int_{p} \left\{ \stackrel{3}{V}_{2,1}-\stackrel{3}{V}_{1,2} \right\} d l_{\rm E}
\right] \nonumber \\
& - & \int_{\rm l.o.s.}
\left(
\left\{ \stackrel{3}{V}_{3,2}-\stackrel{3}{V}_{2,3} \right\}\hat{k}_{(1)}^1 +
\left\{ \stackrel{3}{V}_{1,3}-\stackrel{3}{V}_{3,1} \right\}\hat{k}_{(1)}^2
\right) d l_{\rm E} \nonumber
\\
& -&  \frac{2}{c^2} \int_{\rm l.o.s.}
\left\{ \phi_{,1} \stackrel{3}{V}_{2}-\phi_{,2} \stackrel{3}{V}_{1} \right\} d l_{\rm E} \nonumber \\
& +& \left.
\int_{\rm l.o.s.}
\left\{ \stackrel{5}{V}_{2,1}-\stackrel{5}{V}_{1,2} \right\} d l_{\rm E}
\right\} + {\cal O} (\varepsilon^7), \nonumber
\end{eqnarray}
where comma stands for differentiation and $p$ is the spatial
projection of the null geodesics. The first term in the right hand
side of Eq.~(\ref{far8}) gives the main contribution to the
gravitational Faraday rotation \cite{io04far}. Other terms represent
higher-order corrections. The second and the third contributions
account for the integration along the deflected path $p$. The fourth
term in Eq.~(\ref{far8}) derives from the difference between the
gravito-magnetic potentials ${\bf A}$ and ${\bf V}$.

\section{The thin lens approximation}
\label{thin}

In almost all astrophysical systems, the gravitational lens can be
treated as geometrically thin. The extent of the lens in the direction
of the incoming ray is small compared to both the distances between
lens and observer and lens and source, so that the maximal deviation
of the actual ray from a light path propagating through unperturbed
space-time is small compared to the length scale on which the
gravitational field changes. In this picture, gravitational effects
are quite local to the lens vicinity.

The lens plane is transverse to the incoming, unperturbed light ray
direction. The position vector in the lens plane is the spatial vector
$\bfxi \equiv \left\{ x^1,x^2,0 \right\}$. We introduce the coordinate
$l$ along the $x^3$-axis, so that the lens plane corresponds to $l=0$.
The Born approximation assumes that rays of electromagnetic radiation
propagate along straight lines. The integration along the line of
sight (l.o.s.) is accurate enough to evaluate the main contribution to
the Skrotskii effect. To this order, we will employ the unperturbed
Minkowski metric $\eta_{\alpha \beta}=(1,-1,-1,-1)$ for operations on
vectors.

Let us define the weighted average of a quantity ${\cal Q}$ along the
line of sight,
\begin{equation}
\lab{wf12}
\langle {\cal Q} \rangle_{\rm l.o.s.} (\bfxi)\equiv \frac{\int {\cal Q} \ \rho(\bfxi ,l)\ dl}{\Sigma(\bfxi)}.
\end{equation}
where $\Sigma$ is the surface mass density of the deflector,
\begin{equation}
\lab{wf11}
\Sigma(\bfxi)\equiv \int \rho( \bfxi,l )\ dl.
\end{equation}

The leading contribution to the Skrotskii effect can be written as
\cite{io04far}
\begin{equation}
\int _{\rm l.o.s.} \left. \nabla {\times} \stackrel{3}{\bf V}
\right|_{\rm l.o.s.} d l_{\rm E}.
\end{equation}
Inserting the expression for the gravito-magnetic potential and
performing the integration along the line of sight, we get
\begin{equation}
\label{lea1}
\Omega_{\rm Sk} = \frac{4 \mu}{c^3} \int \frac{ \Sigma(\bfxi^{'})
\langle L_{\rm l.o.s.} \rangle_{\rm l.o.s.} (\bfxi^{'},\bfxi) }{\left|
\bfxi -\bfxi^{'} \right|} d^2 \xi^{'}
\end{equation}
where $L_{\rm l.o.s.}$ is the angular momentum, per unit mass, of a
mass element at $\bfxi^{'}$ with respect a photon with impact
parameter $\bfxi$,
\begin{equation}
\langle L_{\rm l.o.s.} \rangle_{\rm l.o.s.} = \langle  v^1 \rangle_{\rm l.o.s.}
(\xi^2 -\xi^{2'}) - \langle  v^2 \rangle_{\rm l.o.s.} (\xi^1
-\xi^{1'} );
\end{equation}
$v^1$ and $v^2$ are the components of $\bf v$ along
 the $\xi^1$- and the $\xi^2$-axes, respectively.

The angular momentum of the lens can determine a direction of rotation
for the polarization vector. When its projection along the line of
sight is not null, a handness is introduced in the system and some
rotation can be caused. Once the deflector is oriented such that
$\langle L_{\rm l.o.s.}\rangle_{\rm l.o.s.}=0$, the photon moves like
in an equatorial plane and a mirror symmetry about the lens plane
holds. Such a symmetry cancels out the effect \cite{dy+sh92}.

\subsection{Spherical lenses}
\label{sphe}

In most of the astrophysical situations, lenses can be approximated
with nearly spherical mass density profiles. Let us consider spherical
deflector in rigid rotation (constant angular velocity $\bfom$), when
${\bf v}= \bfom {\times} \bfx $. We limit to a slow rotation so that the
deformation caused by rotation is negligible and the body has a
spherical symmetry.

Taking the centre of the source as the spatial origin of a background
inertial frame, the weighted average of the rotational velocity along
the line of sight turns out to be
\begin{equation}
\label{sph1}
\langle {\bf v} \rangle_{\rm l.o.s.} = \left\{ \xi^2 \omega^1-\xi^1 \omega^2,
-\xi^2 \omega_{\rm l.o.s.} , \xi^1 \omega_{\rm l.o.s.} \right\}.
\end{equation}
We can, now, evaluate the integral in Eq.(\ref{lea1}); the
polarization vector of a linearly polarized light ray with impact
parameter $|\bfxi|$ is rotated by
\begin{equation}
\label{sph2}
\Omega_{\rm Sk} =
\frac{4 \pi G \mu}{c^3} \omega_{\rm l.o.s.} M_{\rm Cyl}(> |\bfxi|)
\end{equation}
where $M_{\rm Cyl}(>|\bfxi|)$ is the projected lens mass outside
$|\bfxi|$,
\begin{equation}
M_{\rm Cyl}(>|\bfxi|) =2 \pi \int_{|\bfxi|}^{+\infty}
\Sigma(\xi^{'})
\xi^{'} d
\xi^{'}.
\end{equation}
The gravitational Faraday rotation is proportional to the mass outside
$|\bfxi|$ and can be significant for lenses with slowly decreasing
mass density. $\Omega_{\rm Sk}$ can be compared to the contribution to
bending of light due to the gravito-magnetic field, $\Omega_{\rm GRM}$
(see Eqs.~(15,~16) in Sereno \& Cardone~\shortcite{se+ca02}).
Typically, $\Omega_{\rm Sk}$ and $\Omega_{\rm GRM}$ are of the same
order. Contrary to the Faraday effect, the gravito-magnetic
contribution to the deflection angle depends only on the component of
the angular momentum in the lens plane \cite{io02frm}. Whereas the
components of the angular momentum along the $\xi^i$-axis, related to
the momentum of inertia of the mass within $\bxi$ about a central
axis, contribute only to $\Omega_{\rm GRM}$ \cite{se+ca02}, the
contribution due to the mass outside the impact parameter appears in
both $\Omega_{\rm Sk}$ and $\Omega_{\rm GRM}$.

Let us consider an isothermal sphere (IS). The IS is a density profile
widely used to model systems on very different scales, from galaxy
haloes to clusters of galaxies. The surface mass density is
\begin{equation}
\Sigma^{\rm IS} =\frac{\sigma^2_{\rm v}}{2G}\frac{1}{\sqrt{|\bfxi|^2+\xi_{\rm c}^2}},
\end{equation}
where $\sigma_{\rm v}$ is the velocity dispersion and $\xi_{\rm c}$ is
a finite core radius. Since the total mass is divergent, we introduce
a cut-off radius $R \gg |\bfxi|$. The gravitational Faraday rotation is
\begin{equation}
\Omega_{\rm Sk}^{\rm IS} (|\bfxi|) =
\frac{8 \pi G \mu}{c^3} \sigma^2_{\rm v} \omega_{\rm l.o.s.}
 \left( \sqrt{R^2+\xi_{\rm c}^2} - \sqrt{|\bfxi|^2+\xi_{\rm c}^2}\right)
\end{equation}
In particular, in the inner regions ($|\bfxi| \ll R$), the above
equation reduces to
\begin{equation}
\label{is5}
\Omega_{\rm Sk}^{\rm IS} (|\bfxi| \ll R) =
\frac{8 \pi G \mu}{c^3} \sigma^2_{\rm v} \omega_{\rm l.o.s.} R .
\end{equation}

In the case of lensing of distant quasars by foreground galaxies,
images may form inside the galaxy radius. We can model a typical
galaxy as a singular ($\xi_c=0$) IS with $\sigma_{\rm v} \sim
200$~km~s$^{-1}$, $R
\stackrel{<}{\sim} 20$~Kpc and $J \sim 0.1 M_{\odot}$~Kpc$^2$s$^{-1}$,
as derived from numerical simulations \cite{vit+al01}. From
Eq.~(\ref{is5}), we find $\Omega_{\rm Sk}^{\rm IS} \simeq
6$~milliarcsec.

Let us consider $\Omega_{\rm GRM}^{\rm IS}$ for an isothermal sphere.
It is \cite{se+ca02},
\begin{equation}
\label{is6}
\Omega_{\rm GRM}^{\rm IS} (|\bfxi| \ll R) =
\frac{8 \pi G \mu}{c^3} \sigma^2_{\rm v}
\sqrt{ \left( \omega^1\right)^2 +\left( \omega^2\right)^2 } R .
\end{equation}
$\Omega_{\rm Sk}^{\rm IS}$ and $\Omega_{\rm GRM}^{\rm IS}$ are of the
same order but depend on different components of the angular momentum
of the lens.

\section{Higher order corrections}
\label{high}

To determine higher order terms in the gravitational Faraday rotation,
it is enough to consider only the monopole term in the
gravito-electric field but we must go over the dipole moment in the
gravito-magnetic potential. We want now discuss higher order terms in
a rotating black hole.

\subsection{The Kerr black hole}
\label{kerr}

The Faraday effect in a Kerr metric, in the weak field limit, has been
already discussed by several authors \cite{ish+al88,nou99}. The right
order of approximation in the calculation has been ascertained, but a
disagreement on the numerical value still persists. Now, we want to
apply our formalism to light rays passing through the vacuum region
outside a rotating black hole.

For a source at rest at the origin of the coordinates, assuming that
the polar axis of the coordinate system coincides with the rotation
axis, the weak field limit of the Kerr metric in Boyer-Lindquist
coordinates reads
\begin{equation}
\label{ker1}
ds^2 \simeq ds^2_{\rm Sch}-\frac{4 \mu G J}{c^3 r^2}\sin^2 \theta (r d
\phi) (c dt),
\end{equation}
where $ds^2_{\rm Sch}$ is the line element for a spherically symmetric
solution and where terms of quadratic and higher order in the angular
momentum $J$ have been neglected.

The metric in Eq.~(\ref{ker1}) can be expressed in an equivalent
isotropic form by introducing a new radius variable, $\rho$, such as
\begin{equation}
\label{ker2}
r \simeq \rho \left( 1+ \frac{G M}{c^2 \rho}  +{\cal O}(\varepsilon^4)
\right),
\end{equation}
where $M$ is the total mass of the black hole. Substituting
Eq.~(\ref{ker2}) in Eq.~(\ref{ker1}) and introducing quasi-Minkovskian
coordinates related to $(\rho, \theta,
\phi)$ by the usual transformation rules, we get the weak field limit
of the Kerr metric in the isotropic form. By arbitrarily orientating
the angular momentum ${\bf J}$, we get
\begin{eqnarray}
\label{rn4}
ds^2  & \simeq  & \left( 1-\frac{2G M}{c^2 x} \right) \left( c dt
\right)^2
-\left( 1+\frac{2 \gamma G M}{c^2 x} \right) \delta_{ij} d x^i d x^j \nonumber
\\
& -& \frac{8\mu}{c^3} \left( V_i dx^i \right)\left( c dt \right)
\end{eqnarray}
where $x=\sqrt{\delta_{ij} x^i x^j }$ and
\begin{eqnarray}
{\bf V} &  \simeq & \stackrel{3}{\bf V} + \stackrel{5}{\bf V}
\nonumber \\
\stackrel{3}{\bf V} & = & - \frac{G}{2} \frac{ {\bf J} {\times} \bfx }{x^3},
\label{ker3} \\
\stackrel{5}{\bf V} & = & - \stackrel{3}{\bf V}
\left( \frac{G M}{c^2} \frac{1}{x}\right). \label{ker4}
\end{eqnarray}
In the above equations and in what follows, operations on vectors are
performed using $\eta_{\alpha \beta}$. Eqs.~(\ref{ker3},~\ref{ker4})
are also valid for large distances from an isolated physical system.
Since,
\begin{equation}
\nabla {\times} \stackrel{3}{\bf V}( \bfx) = \frac{G}{2} \frac{{\bf J} -3( {\bf J} {\cdot} \hat{\bf x} )
\hat{\bf x} }{x^3} ,
\end{equation}
it is easy to verify that, to order ${\cal O} (\varepsilon^3)$, there
is no gravitational Faraday rotation \cite{ple60,ko+ma02,god70}. Let
us consider higher order terms.

The unit wave-vector, to order ${\cal O}(\varepsilon^2)$ reads
\begin{eqnarray}
\label{wav1}
\hat{\bf k}_{(1)} & =& (1+\gamma)
\left\{
-\frac{R_{\rm Sch}}{2} \left[ 1+\frac{l}{\sqrt{l^2+|\bfxi|^2} }\right] \frac{\xi^1}{|\bfxi|^2},
\right. \nonumber \\
& -& \left. \frac{R_{\rm Sch}}{2}\left[
1+\frac{l}{\sqrt{l^2+|\bfxi|^2}}\right]
\frac{\xi^2}{|\bfxi|^2}, \frac{\phi}{c^2}
\right\},
\end{eqnarray}
where $R_{\rm Sch} \equiv 2GM/c^2$ is the Schwarzschild radius of the
lens. The deflected path to order $G$ is actually the path in a
spherically symmetric metric. To this order, the photon path lies in
the plane through source, lens and observer.

After lengthy but straightforward calculations, we can perform the
integrations in Eq.~(\ref{far8}). We find\footnote{Our $J_{\rm
l.o.s.}$ corresponds to $-a m \cos \vartheta_0$ in Ishihara et al.
~\shortcite{ish+al88} and Nouri-Zonoz~\shortcite{nou99}.}
\begin{equation}
\label{ker6}
\Omega_{\rm Sk}^{\rm Kerr} \simeq -\mu \frac{\pi}{4} \frac{G^2 M}{c^5} \frac{J_{\rm l.o.s.}}{\xi^3} + {\cal O}(\varepsilon^7) .
\end{equation}
As already known \cite{ish+al88,nou99}, the net rotation is of order
${\cal O}(\varepsilon^5)$.

Only the projection of the angular momentum along the line-of-sight
enters the effect. In Nouri-Zonoz~\shortcite{nou99}, the integration
is performed along the unperturbed path. However, contributions from
the deflected path cancel out. The parameter $\gamma$, which measures
space curvature produced by mass, does not enter the effect.

\section{Conclusions}
\label{conc}

We have discussed the theory of the gravitational Faraday rotation in
the weak field limit in a metric theory of gravity. Due to a
gravito-magnetic field, a net rotation of the polarization plane can
occur. Unlike the phenomena of bending of light and time delay of
electromagnetic waves, gravito-magnetism appears in the leading term
of the rotation angle of the plane of polarization.

Useful formulae have been derived for relevant astrophysical systems.
By applying the thin lens approximation, we obtained quite compact
expressions for extended spinning gravitational lenses. The
gravitational Faraday increases for lenses with a slowly decreasing
mass density. The effect of a gravito-magnetic field is of the same
order in both bending of light and rotation of the plane of
polarization. Contrary to the gravito-magnetic deflection, the
Skrotskii effect depends only on the component of the angular momentum
along the line of sight. Furthermore, whereas the rotation of the
deflector can break the cylindrical symmetry of the deflection angle
in the lens plane, no azimuthal dependence enters the gravitational
Faraday rotation. A non null projection of the total angular momentum
of the lens along the line of sight introduces a handness in the
system but no preferred direction in the alignment of the light beam.

To consider higher order effects, we have considered light rays
propagating outside a rotating black hole. Due to the order of
approximation, the effect on light rays propagating in the vacuum
region outside the event horizon of a Kerr black hole has been often
missed. Applying a method based on the use of the Walker-Penrose
constant up to higher-order terms, Ishihara et
al.~\shortcite{ish+al88} first calculated the angle of rotation due to
the presence of the black hole's spin. Nouri-Zonoz~\shortcite{nou99}
used the 1+3 formulation of stationary space-times and the
quasi-Maxwell form of the vacuum Einstein equations to study Kerr
spaces. The results in Ishihara et al.~\shortcite{ish+al88} and
Nouri-Zonoz~\shortcite{nou99} are of the same order but differ for a
numerical factor. Here, we have properly considered all the
contributions to the gravitational Faraday rotation, In particular,
the effect of the curved path has been accounted for. The
gravitational Faraday rotation depends linearly on $\mu$, the
parameter which quantify and measures the gravito-magnetic effect.
Even at higher order of approximation, $\mu$ is the only parameter
which enters the effects. Other parametrized post-Newtonian
parameters, such as $\gamma$, do not appear. For $\mu
=1$, our result agrees with Nouri-Zonoz~\shortcite{nou99}.

Techniques involving the angular relationship between the intrinsic
polarization vectors and the morphological structure of extended radio
jets have been proposed to probe the mass distribution of foreground
lensing galaxies \cite{kro+al91,kro+al96}. The method has been then
extended in Surpi \& Harari~\shortcite{su+ha99} to consider the effect
of weak-lensing by large scale structure. The basic principle
underlying these investigations is that the polarization vector of the
radiation would not be rotated by the gravitational field of the
deflector.

Here, we have considered gravitational Faraday rotation by
astrophysically significant lens models, such as an isothermal sphere.
The Skrotskii effect turns out to be negligible with respect to today
observational uncertainties. Anyway, high quality data in total flux
density, percentage polarization and polarization position angle at
radio frequencies already exist for multiple images of some
gravitational lensing systems, like B0218+357 \cite{big+al99}. Since
the impressive development of technological capabilities, it is not a
far hypothesis to obtain observational evidences in a next future.

Furthermore, polarization measurements of the cosmic microwave
background radiation should probe very subtle effects, such as
primordial gravity wave signals or a remnant from the reionization
epoch. A full understanding of the whole foreground contamination
turns out to be crucial in data analysis. So, a proper modeling and
quantification of the gravitational Faraday effect is necessary.


\begin{thebibliography}{99}

\bibitem[\protect\citename{Balazs }1958]{bal58}
Balazs, N.L., 1958, Phys. Rev. 110, 236

\bibitem[\protect\citename{Biggs et al. }1999]{big+al99}
Biggs, A.D., Browne, I.W.A., Helbig, P., Koopmans, L.V.E., Wilkinson,
P.N., Perley, R.A., 1999, MNRAS, 304, 349

\bibitem[\protect\citename{Ciufolini \& Pavlis }1998]{ciu+al98}
Ciufolini, I., Pavlis, E.C., 1998, Science, 279, 2100

\bibitem[\protect\citename{Ciufolini \& Wheeler }1995]{ci+wh95}
Ciufolini I., Wheeler J.A., 1995, Gravitation and Inertia, Princeton
University Press, Princeton

\bibitem[\protect\citename{Connors et al. }1980]{con+al80}
Connors, P.A., Piran, T., Stark, R.F., 1980, ApJ, 235, 224

\bibitem[\protect\citename{Dyer \& Shaver }1992]{dy+sh92}
Dyer, C.C., Shaver, E.G., 1992, ApJ, 390, L5

\bibitem[\protect\citename{Fishbach \& Freeman }1980]{fi+fr80}
Fishbach E., Freeman B.S., 1980, Phys. Rev. D 22, 2950

\bibitem[\protect\citename{Godfrey }1970]{god70}
Godfrey, B.B., 1970, Phys. Rev. D, 1, 2721

\bibitem[\protect\citename{Ishihara et al. }1988]{ish+al88}
Ishihara, H., Takahashi, M., Tomimatsu, A., 1988, Phys. Rev. D, 38,
472

\bibitem[\protect\citename{Kopeikin \& Mashoon }2002]{ko+ma02}
Kopeikin, S., Mashoon, B., 2002, Phys. Rev. D, 65, 064025

\bibitem[\protect\citename{Kronberg }1991]{kro+al91}
Kronberg, P.P., Dyer, C.C., Burbidge, E.M., Junkkarinen, 1991, ApJ,
367, L1

\bibitem[\protect\citename{Kronberg et al. }1996]{kro+al96}
Kronberg, P.P., Dyer, C.C., R\"{o}ser, H.-J., 1996, ApJ, 472, 115

\bibitem[\protect\citename{Landau \& Lifshits }1985]{la+li85}
Landau L.D., Lifshits E.M., 1985, Teoria dei Campi, Editori Riuniti,
Roma

\bibitem[\protect\citename{Matzner et al. }1981]{ma+ri81}
Matzner R.A., Richter G.W., 1981, Astrophys. Space Sci., 79, 119

\bibitem[\protect\citename{Misner et al. }1973]{mtw73}
Misner, C., Thorne, K.S., Wheeler, J.A., 1973, Gravitation, Freeman,
San Francisco

\bibitem[\protect\citename{Nouri-Zonoz }1999]{nou99}
Nouri-Zonoz, M., 1999, Phys. Rev. D, 60, 024013

\bibitem[\protect\citename{Piran \& Safier }1985]{pi+sa85}
Piran, T., Safier, P.N., 1985, Nat, 318,271

\bibitem[\protect\citename{Petters et al. }2001]{pet+al01}
Petters A.O., Levine H., Wambsganss J., 2001, Singularity Theory and
Gravitational Lensing, Birkh\"{a}user, Boston

\bibitem[\protect\citename{Plebanski }1960]{ple60}
Plebanski, J., 1960, Phys. Rev, 118, 1396

\bibitem[\protect\citename{Richter \& Matzner }1982]{ri+ma82a}
Richter G.W., Matzner R.A., 1982, Phys. Rev. D, 26, 1219

\bibitem[\protect\citename{Rytov }1938]{ryt38}
Rytov, S.M., 1938, (Dokl.) Acad. Sci. URSS, 18, 263

\bibitem[\protect\citename{Schneider et al. }1992]{sef}
Schneider P., J. Ehlers J., Falco E.E., 1992, Gravitational Lenses,
(Springer, Berlin)

\bibitem[\protect\citename{Sereno }2002]{io02frm}
Sereno, M., 2002, Phys. Lett. A., 305, 7; [astro-ph/0209148].

\bibitem[\protect\citename{Sereno }2003a]{io03ppn}
Sereno, M., 2003a, Phys. Rev. D, 67, 064007; [astro-ph/0301290].

\bibitem[\protect\citename{Sereno }2003b]{io03hom}
Sereno, M., 2003b, MNRAS, 344, 942; [astro-ph/0307243]

\bibitem[\protect\citename{Sereno }2004]{io04far}
Sereno, M., 2004, Phys. Rev. D, 69, 087501; [astro-ph/0401295]

\bibitem[\protect\citename{Sereno \& Cardone }2002]{se+ca02}
Sereno, M., Cardone, V.F., 2002, A\&A, 396, 393; [astro-ph/0209297].

\bibitem[\protect\citename{Skrotskii }1957]{skr57}
Skrotskii, G.B., 1957, Dokl. Akad. Nauk USS, 114, 73

\bibitem[\protect\citename{Su \& Mallett }1980]{su+ma80}
Su, F.S.O., Mallett, R.L., 1980, ApJ, 238, 1111

\bibitem[\protect\citename{Surpi \& Harari }1999]{su+ha99}
Surpi, G.C., Harari, D.D., 1999, ApJ, 515, 455

\bibitem[\protect\citename{Vitviska et al. }2002]{vit+al01}
Vitviska M., Klypin A., Kravtsov A.V., Wechsler, R.H., Primack J.R.,
Bullock J.S., 2002, ApJ  581 799

\bibitem[\protect\citename{Will }1993]{wil93}
Will, C.M., 1993, Theory and Experiment in Gravitational Physics, rev.
ed., Cambridge University Press, Cambridge

\end{thebibliography}
\end{document}